\documentclass[12pt,epsf]{article} 
\usepackage{graphicx,epsfig}
\def\simg{{\ \lower-1.2pt\vbox{\hbox{\rlap{$>$}\lower6pt\vbox{\hbox{$\sim$}}}}\ }}
\def\siml{{\ \lower-1.2pt\vbox{\hbox{\rlap{$<$}\lower6pt\vbox{\hbox{$\sim$}}}}\ }}

\def\als{\alpha_{s}}
\def\al{\alpha}
\def\lQ{\Lambda_{\rm QCD}}

\def\dsl{\,\raise.15ex\hbox{/}\mkern-13.5mu D}

\newcommand{\nn}{\nonumber}
\newcommand{\be}{\begin{equation}}
\newcommand{\ee}{\end{equation}}
\newcommand{\bea}{\begin{eqnarray}}
\newcommand{\eea}{\end{eqnarray}}

\newcommand{\Appendix}[1]%
    {%
     \section{#1}%
      }

\begin{document}\setlength{\unitlength}{1mm}

\begin{titlepage}
\begin{flushright}
\tt{UB-ECM-PF-04-21}
\end{flushright}

\vspace{1cm}
\begin{center}
\begin{Large}
{\bf The Renormalization group in non-relativistic theories}\\[2cm]
\end{Large} 
{\large Antonio Pineda}\footnote{This work was supported in part by
MCyT and Feder, FPA2001-3598, by CIRIT, 2001SGR-00065
and by the EU network EURIDICE, HPRN-CT2002-00311.
}\\
{\it Dept. d'Estructura i Constituents de la Mat\`eria, 
     U. Barcelona \\ Diagonal 647, E-08028 Barcelona, Catalonia, Spain}\\
\end{center}

\vspace{1cm}

\begin{abstract}
The resummation of logarithms in Quantum Field Theories 
is a long tale plenty of successes, yet the resummation 
of logarithms in non-relativistic theories has remained 
elusive. This was the most frustrating, since the first 
quantum field theory log ever computed was the Lamb shift one. 
We briefly review recent progress on the resummation of logarithms of 
$\alpha$, which appear in the physics of non-relativistic 
states, using effective field theories. 
We put special emphasis on the basic formalism.  
\vspace{5mm} 
\\ 
PACS numbers: 12.38.Cy, 12.38.Bx, 12.39.Hg, 11.10.St
\end{abstract}

\end{titlepage}
\vfill
\setcounter{footnote}{0} 
\vspace{1cm}

Heavy quark-antiquark systems near threshold are characterized by the
small relative velocity $v$ of the heavy quarks in their center of
mass frame. This small parameter produces a hierarchy of widely
separated scales: $m$ (hard), $mv$ (soft), $mv^2$ (ultrasoft), ...
The factorization between them is efficiently achieved by using
effective field theories, where one can organize the calculation as
various perturbative expansions on the ratio of the different scales,
effectively producing an expansion in $v$. The terms in these series
get multiplied by parametrically large logs: $\ln v$, which can also
be understood as the ratio of the different scales appearing in the
physical system. Again, effective field theories are very efficient in
the resummation of these large logs once a renormalization group (RG)
analysis of them has been performed. Therefore, the use of 
effective field theories has opened the possibility to solve a 
problem, otherwise open, since the first non-relativistic logs 
appeared around fifty years ago (in the Lamb shift in the 
Hydrogen atom). We will review in this paper 
the recent progress achieved in potential NRQCD (pNRQCD) \cite{pNRQCD}
 on the issue of the resummation 
of $\ln v$ terms in the weak coupling regime (we will obviate any 
non-perturbative effects in what follows). In particular, 
we will focus on the theoretical aspects.

Potential NRQCD (pNRQCD) is defined by its particle content and cut-off
$\nu_{pNR}=\{\nu_p,\nu_{us}\}$, where $\nu_p$ is the cut-off of the
relative three-momentum of the heavy quarks and $\nu_{us}$ is the cut-off of the
three-momentum of the gluons and light quarks. They satisfy the following 
inequalities: $|{\bf p}| \ll \nu_p \ll m$ and ${\bf p}^2/m \ll
\nu_{us} \ll |{\bf p}|$, where typically $|{\bf p}| \sim mv$. 
Note that no gluons or light quarks with
momentum of $O(|{\bf p}|)$ are kept dynamical in pNRQCD. The motivation to integrate 
out these degrees of freedom is that they do not appear
as physical (on-shell) states near threshold. Nevertheless, they can appear off-shell 
and, since their momentum is of the order of the relative three-momentum of the heavy 
quarks, integrating them out produces non-local terms (potentials) in three-momentum
space. Indeed, these potentials encode the non-analytical behavior in the transfer
momentum of the heavy quark, ${\bf k}={\bf p}-{\bf p}'$, of the order
of the relative three-momentum of the heavy quarks.

In this paper, we will mainly consider the situation $v \sim \als$. It should be 
clear however that pNRQCD (in the weak coupling regime) is also valid in the 
situation $mv^2 \gg m\als^2$ as long as $v$ is an small parameter. Nevertheless, the 
Coulomb resummation is not necessary in this case.

Formally, we can write the pNRQCD Lagrangian as an expansion in
$1/r$($=1/r$, $p$) and
$1/m$ in the following way:
\be
{\cal L}_{\rm pNRQCD} =\sum_{n=-1}^{\infty}r^n{\tilde V}_n^{(B)}O_n^{(B)}
+{1 \over m}\sum_{n=-2}^{\infty}r^n{\tilde V}_n^{(B,1)}O_n^{(B,1)}
+O\left({1 \over m^2}\right)
, 
\ee
where the above operators and matching coefficients 
should be understood as bare. As
for the renormalized quantities, we define $V$ as the potentials and ${\tilde V}$ as the (almost) dimensionless
constants in it. The latter are in charge of absorbing the divergences
of the effective field theory. Therefore, they will depend on $\nu_p$
and $\nu_{us}$. At next-to-leading order in the multipole expansion the 
Lagrangian can be written as\footnote{$S$ and $O$ stand for 
the heavy quarkonium bilinear 
field in a singlet and octet configuration under ultrasoft 
gauge transformations. They both 
depend on the relative, ${\bf r}$, and center of mass, ${\bf R}$, coordinate.}
\bea
&& L_{pNRQCD} =
\int d^3r d^3 R tr \Biggl\{
 S^{\dagger}
              \Bigl\{
i\partial_0 - h_s
\Bigr\} S
+ O^{\dagger}
                \Bigl\{
iD_0 - h_o
% +{\vec D_{\vec X}^2\over 4m}
\Bigr\} O 
\\
&&
\nonumber
+g V_A{\bf r} \cdot  O {\bf E} ({\bf R} , t)
S^{\dagger}
+g V_A {\bf r} \cdot O^{\dagger} {\bf E} ({\bf R} , t)
S
\\
&&
\nonumber
+{g\over 2} V_B{\bf r}\cdot  O O^{\dagger} {\bf E} ({\bf R}, t)
+{g\over 2} V_B{\bf r} \cdot O^{\dagger} O {\bf E} ({\bf R}, t) \Biggr\}
\,,
\eea
where $h_s$ and $h_o$ are the singlet and octet quantum mechanical 
Hamiltonian. 
For illustration, at low orders (see Ref. \cite{RGmass} for notation 
and further details),
\bea
h_s&=&c_k{{\bf p}^2 \over m}-c_4{{\bf p}^4 \over 4m^3}- C_f {\alpha_{V_{s}} \over r}-{C_fC_A
D^{(1)}_s \over 2mr^2}
\nn
\\
&&- { C_f D^{(2)}_{1,s} \over 2 m^2} \left\{ {1 \over r},{\bf p}^2 \right\}
+ { C_f D^{(2)}_{2,s} \over 2 m^2}{1 \over r^3}{\bf L}^2
+ {\pi C_f D^{(2)}_{d,s} \over m^2}\delta^{(3)}({\bf r})
\nn
\\
& & + {4\pi C_f D^{(2)}_{S^2,s} \over 3m^2}{\bf S}^2 \delta^{(3)}({\bf r})
+ { 3 C_f D^{(2)}_{LS,s} \over 2 m^2}{1 \over r^3}{\bf L} \cdot {\bf S}
+ { C_f D^{(2)}_{S_{12},s} \over 4 m^2}{1 \over r^3}S_{12}(\hat{\bf r})
\,,
\eea
in the equal mass case, where $C_f=(N_c^2-1)/(2N_c)$ and we will set $c_k=c_4=1$. A similar 
expression holds for the octet Hamiltonian changing the label of the 
matching coefficients ($s \rightarrow o$).

The matching process, which basically means the computation of the
potentials, is carried out for a given external incoming (outcoming)
momentum ${\bf p}$ (${\bf p}'$). Therefore, one has to sum over all of them in
the pNRQCD Lagrangian, since they are still physical degrees of
freedom as far as their momentum is below $\nu_p$. In position space, 
this means that an integral over ${\bf x}$, the relative distance
between the heavy quarks, appears in the Lagrangian when written in
terms of the heavy quark-antiquark bilinear field.

Within pNRQCD, integrals over ${\bf p}$ (or ${\bf x}$) appear when
solving the Schr\"o-\\
dinger equation that dictates the dynamics of the
heavy quarkonium near threshold. At low orders, these integrals are
finite effectively replacing ${\bf p}$ by $\sim m\als$. Nevertheless,
at higher orders in quantum mechanics perturbation theory and/or if
some singular enough operators are introduced (as it is the case
of the heavy quarkonium production currents) singularities
proportional to $\ln\nu_p$ appear. These must be absorbed by the potentials
or by the matching coefficients of the currents. 

Let us now describe the matching between QCD and pNRQCD within an RG
framework. We first address the procedure that gives the running of the potentials.
One first does the matching from QCD to NRQCD. The latter depends on
some matching coefficients: $c(\nu_s)$ and $d(\nu_p,\nu_s)$, which can
be obtained order by order in $\als$ (with $\nu_p=\nu_s$) following the
procedure described in Ref. \cite{Manohar} for the $c(\nu_s)$ and 
\cite{Match} for the $d(\nu_p,\nu_s)$. $\nu_s$ is the 
ultraviolet cutoff of the three-momentum of the gluons in NRQCD. 
The $c(\nu_s)$ stand for the
coefficients of the operators that already exist in the theory with only
one heavy quark (ie. HQET) and the $d(\nu_p,\nu_s)$ stand for the
coefficients of the four heavy fermion operators. The starting point of
the renormalization group equation can be obtained from these
calculations by setting $\nu_p=\nu_s=m$ (up to a constant of order
one). In principle, we should now compute the running of $\nu_p$ and
$\nu_s$. The running of the $c(\nu_s)$ can be obtained using HQET
techniques \cite{HQET}. The running of the $d(\nu_p,\nu_s)$ is more
complicated. At one-loop, $\nu_p$ does not appear and we effectively
have $d(\nu_p,\nu_s)\simeq d(\nu_s)$, whose running can also be obtained
using HQET-like techniques \cite{RGmass}. At higher orders, the
dependence on $\nu_p$ appears and the running of the $d(\nu_p,\nu_s)$
becomes more complicated. Fortunately, we need not compute the
running of $d$ in this more general case because, as we will see, the
relevant running of the $d$ for near threshold observables can be
obtained within pNRQCD.

The next step is the matching from NRQCD to pNRQCD. The latter depends
on some matching coefficients (potentials). They typically have the
following structure: ${\tilde V}(c(\nu_s), d(\nu_p,\nu_s),\nu_s,
\nu_{us},r)$. After matching, any dependence on $\nu_s$ disappears since
the potentials have to be independent of $\nu_s$. Therefore, they could
be formally written as ${\tilde V}(c(1/r), d(\nu_p,1/r),
1/r,\nu_{us},r)$. These potentials can be obtained order by order in
$\als$ following the procedure of Refs. \cite{pNRQCD,pNRQED}. 
The integrals in
the matching calculation would depend on a factorization scale $\mu$,
which should correspond either to $\nu_s$ or to $\nu_{us}$. In the explicit calculation, 
they could be distinguished by knowing the UV and infrared (IR) behavior
of the diagrams: UV divergences are proportional to $\ln\nu_s$, which
should be such as to cancel the $\nu_s$ scale dependence inherited from
the NRQCD matching coefficients, and IR divergences to $\nu_{us}$. In
practice, however, as far as we only want to perform a matching
calculation at some given scale $\mu=\nu_s=\nu_{us}$, it is not
necessary to distinguish between $\nu_s$ and $\nu_{us}$ (or if working
order by order in $\als$ without attempting any log resummation).

\begin{figure}
\makebox[0.0cm]{\phantom b}
\put(42,9){$\underbrace{\hbox{~~~~~~~~~~~~~~~~~~~~~~~~~~~~~}}$}
\put(45,2){$1/(E-V^{(0)}_o-{\bf p}^2/m)$}
\put(27,10){\epsfxsize=7truecm \epsfbox{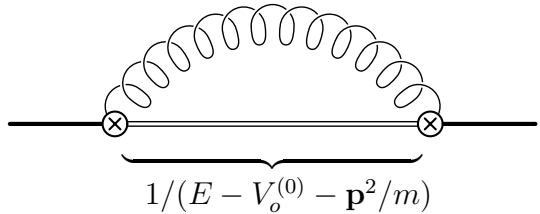}}
\caption{The UV divergences of this diagram in pNRQCD give the 
leading non-trivial ultrasoft running of $\alpha_{V_s}$, 
$D_s^{(1)}$, $D_s^{(2)}$.}
\vspace{3mm}
\label{figus}
\end{figure}

Before going into the rigorous procedure to obtain the RG equations of
the potentials, let us first discuss their structure on physical
grounds. The potential is independent of
$\nu_s$. This allows us to fix $\nu_s$ to $1/r$ that, in a way, could be understood as the
matching scale for $\nu_s$\footnote{In practice, the potential is often
first obtained in momentum space so that one could then set
$\nu_s=k$. Note, however, that this is not equivalent to fix $\nu_s=1/r$, since
finite pieces will appear after performing the Fourier
transform.}. Therefore, $1/r$, the point where the multipole expansion
starts, would also provide with the starting point of the
renormalization group evolution of $\nu_{us}$ (up to a constant of order
one).  The running of $\nu_{us}$ can then be obtained following the
procedure described in Refs. \cite{RG,RGmass}. 
Formally, the renormalization group equations of the matching
coefficients due to the $\nu_{us}$-dependence read 
\be
\nu_{us} {d \over d \nu_{us}}{\tilde V}=B_{{\tilde V}}({\tilde V})
.
\ee
From a practical point of view one can organize the RG equations
within an expansion in $1/m$.
At $O(1/m^0)$, the analysis corresponds to the study of the static
limit of pNRQCD, which has been carried out in
Ref. \cite{RG}. Since ${\tilde V}_{-1}\not= 0$, there are
relevant operators (super-renormalizable terms) in the Lagrangian
and the US RG equations lose the triangular structure that we enjoyed for the
RG equations of $\nu_s$. Still,
if ${\tilde V}_{-1} \ll 1$, a perturbative calculation of the renormalization
group equations can be achieved as a double expansion in ${\tilde V}_{-1}$ and
${\tilde V}_0$, where the latter corresponds to the marginal operators (renormalizable interactions)). 
At short distances ($1/r \gg \lQ$), the static limit of pNRQCD lives in this situation. Specifically, we
have ${\tilde V}_{-1}=\{\al_{V_s},\al_{V_o}\}$, that fulfills ${\tilde
V}_{-1} \sim \als(r) \ll 1$;
${\tilde V}_0=\als(\nu_{us})$  and ${\tilde V}_{1}=\{V_A,V_B\} \sim 1$. Therefore, 
we can calculate the anomalous dimensions order by
order in $\als(\nu_{us})$. In addition, we also have an expansion in
${\tilde V}_{-1}$. Moreover, the specific form of the pNRQCD
Lagrangian severely constrains the RG equations general structure. Therefore,
for instance, the leading non-trivial RG equation of $\al_{V_s}$ reads
\be
\nu_{us} {d\over d\nu_{us}}\al_{V_{s}}
=
{2 \over 3}{\alpha_{s}\over
  \pi}V_A^2\left( \left({C_A \over 2} -C_f\right)\al_{V_o}+
  C_f\al_{V_s}\right)^3
+O({\tilde V}_{-1}^4{\tilde V}_0,{\tilde V}_0^2{\tilde V}_{-1}^3) 
\,.
\ee 
At higher orders in $1/m$ 
the same considerations than for the static limit apply
here as far as the non-triangularity of the RG equations is
concerned. 
In general, we have the structure (${\tilde
V}_m^{(0)} \equiv {\tilde V}_m$) 
\be
\nu_{us} {d\over d\nu_{us}}{\tilde V}_{m}^{(n)}
\sim 
\sum_{\{n_i\}\{m_i\}}{\tilde V}_{m_1}^{(n_1)}{\tilde
V}_{m_2}^{(n_2)}\cdots{\tilde V}_{m_j}^{(n_j)}\,, \quad {\rm with}\quad  
\sum_{i=1}^jn_i=n\;,\, \sum_{i=1}^jm_i=m\,,
\ee
and one has to pick up the leading contributions from all the possible
terms. Actually, as far as the NNLL heavy quarkonium mass is concerned,
the relevant ultrasoft running can be obtained by computing the 
diagram displayed in Fig. 
\ref{figus} (see \cite{logs}) (one also has 
to consider the running of $V_A$, which happens to be zero).

At the end of the day, we
would have ${\tilde V}(c(1/r),d(\nu_p,1/r),1/r,\nu_{us},r)$, where the
running on $\nu_{us}$ is known and also the running in $1/r$ if the $d$
is $\nu_p$-independent. So far, the only explicit dependence of the
potential on $\nu_p$ appears in the $d$. Nevertheless, the potential is
also implicitly dependent on the three-momentum of the heavy quarks
through the requirement $1/r \sim {\bf p} \ll \nu_p$, and also through
$\nu_{us}$, since $\nu_{us}$ needs to fulfill ${\bf p}^2/m \ll \nu_{us}
\ll |{\bf p}|$ in order to ensure that only soft degrees of freedom 
have been integrated out for a given $|{\bf p}|$. This latter requirement 
holds if we fix
$\nu_{us}=\nu_p^2/m$ (this constraint tells you how much you can run
down $\nu_{us}$ in the potential before finding the cutoff $\nu_p^2/m$
caused by the cutoff of ${\bf p}$).

Within pNRQCD, the potentials should be introduced in the
Schr\"odinger equation. This means that integrals over the relative
three-momentum of the heavy quarks take place. When these integrals
are finite one has ${\bf p} \sim 1/r \sim m\als$ and ${\bf p}^2/m \sim
m\als^2$. Therefore, one can lower $\nu_{us}$ down to $\sim m\als^2$
reproducing the results obtained in Ref. \cite{RGmass}. In some cases,
in particular in heavy quarkonium creation, the integrals over ${\bf
p}$ are divergent, and the log structure is dictated by the ultraviolet
behavior of ${\bf p}$ and $1/r$. This means that we can not replace
$1/r$ and $\nu_{us}$ by their physical expectation values but rather by
their cutoffs within the integral over ${\bf p}$. Therefore, for the
RG equation of $\nu_p$, the anomalous dimensions will depend (at
leading order) on ${\tilde
V}(c(\nu_p),d(\nu_p,\nu_p),\nu_p,\nu_p^2/m,\nu_p)$\footnote{Roughly speaking, this result 
can be
thought as expanding $\ln r$ around $\ln \nu_p$ in the potential ie.
\bea
\nn 
{\tilde V}(c(1/r),d(\nu_p,1/r),1/r,\nu_p^2/m,r) &\simeq& {\tilde
V}(c(\nu_p),d(\nu_p,\nu_p),\nu_p,\nu_p^2/m,\nu_p)
\\
&&
+\ln(\nu_pr)r{d \over d r} {\tilde V}\bigg|_{1/r=\nu_p} + \cdots 
\,.
\eea 
The $\ln(\nu_pr)$ terms give subleading contributions to the
anomalous dimension when introduced in divergent integrals over ${\bf
p}$. A more
precise discussion would require a full detailed study within
dimensional regularization. An explicit example of this type of corrections appear in the 
computation of the hyperfine splitting of the heavy quarkonium at 
NLL \cite{KPPSS,Penin:2004xi}.} and the running will go from $\nu_p \sim m$
down to $\nu_p \sim m\als$. Note that, at this stage, a single cutoff,
$\nu_p$, exists and the correlation of cutoffs can be seen. The
importance of the idea that the cutoffs of the non-relativistic
effective theory should be correlated was first realized in Ref. 
\cite{LMR}. 

With the above discussion in mind, the matching between NRQCD and
pNRQCD could be thought as follows. One does the matching by computing
the potentials order by order in $\als$ at the matching scale
$\nu_p=\nu_s=\nu_{us}$ following the procedure of Refs. \cite{pNRQCD,pNRQED} (by
doing the matching at a generic $\nu_p$ some of the running is
trivially obtained).  The structure of the potential at this stage
then reads ${\tilde V}(c(\nu_p),d(\nu_p,\nu_p),\nu_p,\nu_p,\nu_p)$ (and
similarly for the derivatives with respect $\ln r$ of the
potential). This provides the starting point of the renormalization
group evolution of $\nu_{us}$ (up to a constant of order one).  The
running of $\nu_{us}$ can then be obtained following the procedure
described in Refs. \cite{RG,RGmass}. For the final point of the evolution
of $\nu_{us}$, we choose $\nu_{us}=\nu_p^2/m$. At the end of the day, we
obtain ${\tilde V}(c(\nu_p), d(\nu_p,\nu_p),\nu_p,\nu_p^2/m,\nu_p)
\equiv {\tilde V}(\nu_p)$.

The running of $\nu_p$ goes from $\nu_p=m$ (this was fixed when the
matching between QCD and NRQCD was done) up to the physical scale of
the problem $\nu_p \sim m\als$. If the running of the NRQCD
matching coefficients is known, the above result gives the complete running of
the potentials. The procedure to get the running of the $c$ is known
at any finite order. For the $d$ it is just known at one-loop order, 
since, at this order, it is only $\nu_s$-dependent. Nevertheless, at
higher orders, dependence on $\nu_p$ appears. Therefore, the above
method is not complete unless an equation for the running of $\nu_p$ is
provided. This is naturally given within pNRQCD. It appears through the
iteration of potentials. Let us consider this situation more in
detail.  The propagator of the
singlet is (formally)
\be
{1 \over \displaystyle{E-h_s}}
\,.
\ee
At leading order (within an strict expansion in $\als$) the propagator of the singlet reads
%%%%%%%%%%%%%%%%%%%%%%%%%%%%%%%%%%%%%%%%%%%%%%%%%%%%%%%%%%%%%%%%%%%%%%%%%%%%%%%%%%
\begin{figure}[htb]
\makebox[0.5cm]{\phantom b}
\epsfxsize=3truecm \epsfbox{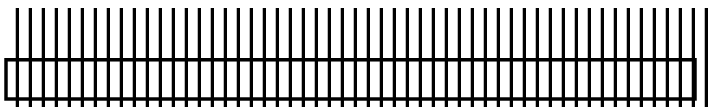}
\put(15,1){$
\label{Gc}
=G_c(E)=\displaystyle{{1 \over \displaystyle{E-h_s^{(0)}}}
=
{1 \over \displaystyle{E-{\bf p}^2/m-C_f \als/ r}}}
\,.$}
\end{figure}
%%%%%%%%%%%%%%%%%%%%%%%%%%%%%%%%%%%%%%%%%%%%%%%%%%%%%%%%%%%%%%%%%%%%%%%%%%%%%%%%%%

If we were interested in computing the spectrum at $O(m\als^6)$, one
should consider the iteration of subleading potentials ($\delta h_s$) in the
propagator as follows:
\be
G_c(E)\delta h_s G_c(E) \cdots \delta h_s G_c (E)
\,.
\ee
In general, if these potentials are singular enough, these
contributions will produce logarithmic divergences due to potential
loops. These divergences can be absorbed in the
matching coefficients, $D^{(2)}_{d,s}$ and $D^{(2)}_{S^2,s}$, of the 
local potentials (those proportional to the $\delta^{(3)}({\bf r}$)) providing
with the renormalization group equations of these matching
coefficients in terms of $\nu_p$. Let us explain how it works in
detail. Since the singular behavior of
the potential loops appears for $|{\bf p}| \gg \als/r$, a perturbative
expansion in $\als$ is allowed in $G_c(E)$, which can be approximated by

%%%%%%%%%%%%%%%%%%%%%%%%%%%%%%%%%%%%%%%%%%%%%%%%%%%%%%%%%%%%%%%%%%%%%%%%%%%%%%%%%%
\begin{figure}[htb]
\makebox[0.5cm]{\phantom b}
\epsfxsize=3truecm \epsfbox{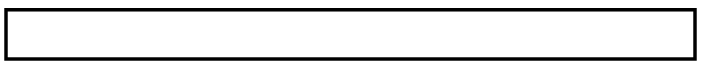}
\put(25,1){$
\label{Gc0}
=G_c^{(0)}(E)=\displaystyle{
{1 \over \displaystyle{E-{\bf p}^2/m}}}
\,.$}
\end{figure}
%%%%%%%%%%%%%%%%%%%%%%%%%%%%%%%%%%%%%%%%%%%%%%%%%%%%%%%%%%%%%%%%%%%%%%%%%%%%%%%%%%

Therefore, a practical 
simplification follows from the fact that the Coulomb potential, $-C_f
{\als \over r}$, can 
be considered to be a perturbation as far as the computation of the
$\ln \nu_p$ ultraviolet 
divergences is concerned. This means that the computation of the
anomalous dimension can be organized within an expansion in
$\als$ and using the free propagators $G_c^{(0)}$. Moreover, each
$G_c^{(0)}$ produces a potential loop and one extra power of $m$ in the
numerator, which kills the powers in $1/m$ of the different
potentials. This allows the mixing of potentials with different powers
in $1/m$. One typical example would be the diagram in
Fig. 2.
 The computation of this diagram would go as follows:
\be
G_c^{(0)}(E)
{\pi C_f D^{(2)}_{d,s} \over m^2}\delta^{(3)}({\bf r})
G_c^{(0)}(E)
C_f {\al_{V_s} \over r}
G_c^{(0)}(E)
{\pi C_f D^{(2)}_{d,s} \over m^2}\delta^{(3)}({\bf r})
G_c^{(0)}(E)
\,.
\ee
Using $\delta^{(3)}({\bf r})=|{\bf r}=0\rangle\langle{\bf r}=0|$, we can see that the 
relevant computation reads (instead of $\al_{V_s}$ one could use
$\als$ since the non-trivial running of $\al_{V_s}$ is a subleading
effect. Nevertheless, we keep $\al_{V_s}$ since it allows to keep track of the
contributions due to the Coulomb potentials)
\bea
&&
\langle{\bf r}=0|
G_c^{(0)}(E)
C_f {\al_{V_s} \over r}
G_c^{(0)}(E)
|{\bf r}=0\rangle
\\
\nn
&&
\qquad
\sim 
\int \frac{
{\rm d}^d p' }{ (2\pi)^d } \int \frac{ {\rm d}^d p }
{ (2\pi)^d } \frac{ m }{{\bf p}'^2 - mE } 
C_f
\frac{ 4\pi\alpha_{V_s} }{ {\bf q}^2 } \frac{ m }{{\bf p}^2-m E } 
\sim
- C_f\frac{m^2\alpha_{V_s}}{16\pi}  
\frac{ 1 }{\epsilon },
\eea
where $D=4+2\epsilon$ and ${\bf q}={\bf p}-{\bf p}'$. This divergence is absorbed in $D_{d,s}^{(2)}$ contributing to its running at
next-to-leading-log (NLL) order as follows 
\be
\label{eqDd}
\nu_p {d \over d\nu_p}D_{d,s}^{(2)}(\nu_p) \sim 
\alpha_{V_s}(\nu_p)D_{d,s}^{(2)2}(\nu_p)+\cdots
\,.
\ee
Therefore, even without knowing the running of the $d$ 
(which need to be known at NLL order in this case), we
can obtain the running of the potential (one can also think of trading
Eq. (\ref{eqDd}) into an equation for $d$, which is the only unknown 
parameter within the potential). This is so because $D_{d,s}^{(2)}$ is 
only needed with LL accuracy in the right-hand side of Eq. (\ref{eqDd}). 

%%%%Figure Ddnup%%%%
\medskip
\begin{figure}[h]
\hspace{-0.1in}
\epsfxsize=2.8in
\centerline{\epsffile{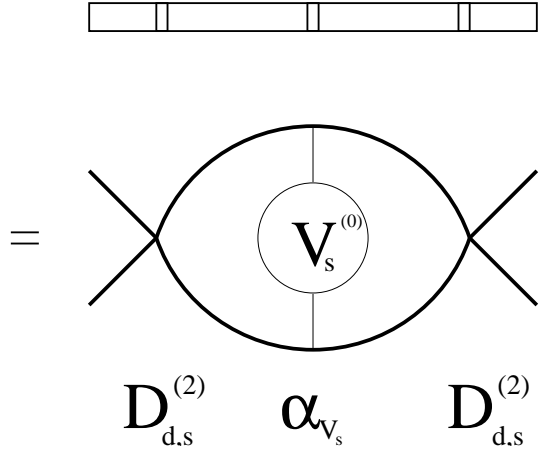}}
\caption {{\it One possible contribution to the running of $D_{d,s}^{(2)}$ at
NLL. The first picture represents the calculation in 
terms of the free quark-antiquark propagator $G_c^{(0)}$ and the 
potentials (the small rectangles). The picture
below is the representation within a more standard diagrammatic
interpretation in terms of quarks and antiquarks. The delta potentials
are displayed as local interactions and the Coulomb potential as an
extended in space (but not in time) object.}}
\label{obs12}
\end{figure}
%%%%End Figure Ddnup%%%%

This finishes the procedure to the RG equations. 
The above method deals with the resummation of logs due to the hard, 
soft and ultrasoft scales. Nevertheless, for some specific 
kinematical situations even smaller scales could appear. Their study, 
however, has not yet been carried out. In any case, pNRQCD can 
be considered to be the right starting point to study these 
kinematical situations. 

This line of investigation has lead to several new results on heavy 
quarkonium physics. They can be summarized in the following way
\begin{itemize}
\item
The correction to the heavy quarkonium energy at NNLL 
\cite{RGmass}, i.e. corrections of order 
\be
\delta E \sim m\alpha^4+m\alpha^5\ln\alpha+m\alpha^6\ln^2\alpha+\cdots\\
\ee
\item
Corrections to the heavy quarkonium hyperfine splitting at LL 
\cite{RGmass} (first obtained in Ref. \cite{HMS}) and 
NLL \cite{KPPSS,Penin:2004xi}
\bea
\delta E_{HF} 
&\sim& m\alpha^4+m\alpha^5\ln\alpha+m\alpha^6\ln^2\alpha+\cdots\\
&+&m\alpha^5+m\alpha^6\ln\alpha+m\alpha^7\ln^2\alpha+m\alpha^8\ln^3\alpha+\cdots
\nn
\eea
\item
The decays are known with NLL accuracy (this result can be easily applied to 
${\bar t}-t$ 
production threshold or non-relativistic sum rules since the running 
of the electromagnetic current matching coefficient is  
the only non-trivial object that appears at NLL running) \cite{csNLL}
\bea
\Gamma(V_Q (nS) \rightarrow e^+e^-) 
&\sim& 
m\alpha^3(1+\alpha^2\ln\alpha+\alpha^3\ln^2\alpha+\cdots)
\\
\nn
\Gamma(P_Q (nS) \rightarrow \gamma\gamma) 
&\sim& 
m\alpha^3(1+\alpha^2\ln\alpha+\alpha^3\ln^2\alpha+\cdots)
\eea
and for the ratio with NNLL accuracy \cite{Penin:2004ay}
\bea
{\Gamma(V_Q (nS) \rightarrow e^+e^-) 
\over
\Gamma(P_Q (nS) \rightarrow \gamma\gamma) }
&\sim&
1+\alpha^2\ln\alpha+\alpha^3\ln^2\alpha+\cdots\\
\nn
&&
\quad +\alpha^3\ln\alpha+\alpha^4\ln^2\alpha+\cdots
\eea
\end{itemize}

There has also been a lot of work on the resummation of logarithms using 
the vNRQCD framework. Unfortunately, its first formulation suffered from 
some mistakes (in particular concerning the treatment of ultrasoft modes), 
which lead to incorrect results for the heavy quarkonium mass at NNLL 
\cite{HMS}
and the electromagnetic current matching coefficient at NLL \cite{v1}. 
Fortunately, they have been solved in Ref. \cite{Hoang:2002yy} and their 
results now agree with those obtained in pNRQCD \cite{RGmass,csNLL}.

These results may have an important 
phenomenological impact in several situations. Let us enumerate a few of 
them. The determination of the bottom and charm masses (using the 
experimental value of the ground state heavy quarkonium masses or 
non-relativistic sum rules). The 
determination of the $\eta_b(1S)$ mass, the hyperfine splitting of 
the ground state $B_c$ system, or theoretical improved determinations of  
the $\eta_c$. One can also try to obtain improved determinations 
for the inclusive electromagnetic decays of the heavy quarkonium. 
On the other hand the application of this program to 
$t$-$\bar t$ production near threshold at the Next Linear Collider ´
is one of the main 
motivations to undergo these computations.  
Moreover, it would also be interesting 
to try to perform the resummation of logarithms in semi-inclusive 
radiative decays \cite{GarciaiTormo:2004jw}.
A review on the phenomenological consequences of these results is 
presented elsewhere.

\medskip
 
\noindent
{\bf Acknowledgments}\\
The author would like to acknowledge very pleasant 
collaborations with Joan Soto, Bernd Kniehl, Sasha Penin, Vladimir Smirnov  
and Matthias Steinhauser on which parts of the work reported 
here are based.

%%%%%%%%%%%%%%%%%%%%% BIBLIOGRAPHY %%%%%%%%%%%%%%%%%%%%%%%%%%%%%%%%%%%%

\end{document}